\documentclass[aps,prl,twocolumn,superscriptaddress,oneside,floatfix,amsmath,showpacs,amssymb]{revtex4}
\usepackage{graphicx}
\usepackage{dcolumn}
\usepackage{bm}

\begin{document}

\newcommand{\Y}{YBa$_2$Cu$_3$O$_y$}
\newcommand{\ie}{\textit{i.e.}}
\newcommand{\eg}{\textit{e.g.}}
\newcommand{\etal}{\textit{et al.}}


\title{de Haas-van Alphen Oscillations in the Underdoped High-Temperature Superconductor YBa$_2$Cu$_3$O$_{6.5}$}


\author{Cyril~Jaudet}
\affiliation{Laboratoire National des Champs Magn\'{e}tiques
Puls\'{e}s (CNRS-UPS-INSA), Toulouse, France}

\author{David~Vignolles} \email{vignolle@lncmp.org}
\affiliation{Laboratoire National des Champs Magn\'{e}tiques
Puls\'{e}s (CNRS-UPS-INSA), Toulouse, France}

\author{Alain~Audouard}
\affiliation{Laboratoire National des Champs Magn\'{e}tiques
Puls\'{e}s (CNRS-UPS-INSA), Toulouse, France}

\author{Julien~Levallois}
\affiliation{Laboratoire National des Champs Magn\'{e}tiques
Puls\'{e}s (CNRS-UPS-INSA), Toulouse, France}

\author{D.~LeBoeuf}
\affiliation{D\'epartement de physique \& RQMP, Universit\'e de
Sherbrooke, Sherbrooke, Canada}

\author{Nicolas~Doiron-Leyraud}
\affiliation{D\'epartement de physique \& RQMP, Universit\'e de Sherbrooke, Sherbrooke, Canada}

\author{B.~Vignolle}
\affiliation{Laboratoire National des Champs Magn\'{e}tiques
Puls\'{e}s (CNRS-UPS-INSA), Toulouse, France}

\author{M.~Nardone}
\affiliation{Laboratoire National des Champs Magn\'{e}tiques Puls\'{e}s (CNRS-UPS-INSA), Toulouse, France}

\author{A.~Zitouni}
\affiliation{Laboratoire National des Champs Magn\'{e}tiques
Puls\'{e}s (CNRS-UPS-INSA), Toulouse, France}

\author{Ruixing~Liang}
\affiliation{Department of Physics and Astronomy, University of British Columbia, Vancouver, Canada}
\affiliation{Canadian Institute for Advanced Research, Toronto, Canada}

\author{D.A.~Bonn}
\affiliation{Department of Physics and Astronomy, University of British Columbia, Vancouver, Canada}
\affiliation{Canadian Institute for Advanced Research, Toronto, Canada}

\author{W.N.~Hardy}
\affiliation{Department of Physics and Astronomy, University of British Columbia, Vancouver, Canada}
\affiliation{Canadian Institute for Advanced Research, Toronto, Canada}

\author{Louis~Taillefer}
\affiliation{D\'epartement de physique \& RQMP, Universit\'e de
Sherbrooke, Sherbrooke, Canada} \affiliation{Canadian Institute
for Advanced Research, Toronto, Canada}

\author{Cyril~Proust} \email{proust@lncmp.org}
\affiliation{Laboratoire National des Champs Magn\'{e}tiques
Puls\'{e}s (CNRS-UPS-INSA), Toulouse, France}

\date{\today}


\begin{abstract}

The de Haas-van Alphen effect was observed in the underdoped
cuprate YBa$_2$Cu$_3$O$_{6.5}$ via a torque technique in pulsed
magnetic fields up to 59~T. Above a field of $\sim$30~T, the
magnetization exhibits clear quantum oscillations with a single
frequency of $540$~T and a cyclotron mass of $1.76$ times the free
electron mass, in excellent agreement with previously observed
Shubnikov-de Haas oscillations. The oscillations obey the standard
Lifshitz-Kosevich formula of Fermi-liquid theory. This
thermodynamic observation of quantum oscillations confirms the
existence of a well-defined, closed, and coherent, Fermi surface
in the pseudogap phase of cuprates.

\end{abstract}


\maketitle


Progress towards a more complete understanding of the high
temperature superconductors involves a few key questions, one of
which is to understand the nature of the pseudogap phase in the
underdoped region of the phase diagram. Angle-resolved
photoemission spectroscopy (ARPES) shows an apparent destruction
of the Fermi surface (FS): below the pseudogap temperature
$T^{\star}$, a gap opens up along the $(\pi,0)$ direction,
producing a set of disconnected Fermi arcs. \cite{Norman98}
Recently, quantum oscillations, a clear signature of a
Fermi-liquid ground state, were observed in the Hall resistance of
YBa$_2$Cu$_3$O$_{6.5}$. \cite{Doiron07} Since quantum oscillations
are a direct consequence of the quantization of closed orbits in a
magnetic field, this result suggests that the FS consists of small
pockets, rather than Fermi arcs. Very recently, Shubnikov-de Haas
oscillations have also been observed in the stoichiometric
compound YBa$_2$Cu$_4$O$_8$, \cite{Bangura07, Yelland07}
suggesting that they are generic to the CuO$_2$ planes rather than
some feature of the band structure specific to
YBa$_2$Cu$_3$O$_{6.5}$. \cite{Carrington07, Bangura07} The
frequency of the oscillations gives the area of the FS, but
neither the location in $\it{k}$-space, nor the number of pockets
are known. A comparison with ARPES measurements (assuming only
half of each pocket is detected) suggests a four nodal pockets
scenario \cite{Julian07}. However, when the density of carriers is
estimated from the Luttinger sum rule, an obvious violation is
found in both systems studied so far. In addition, the four nodal
pockets scenario presumes that these pockets are hole-like, in
contradiction with the negative sign of the Hall effect observed
in both systems in the normal state at low temperature.
\cite{LeBoeuf07} An alternative scenario has been proposed whereby
the FS of these Y-based cuprates consists of both electron and
hole pockets, probably arising from a reconstruction of the FS.
\cite{LeBoeuf07, Chakravarty08} Moreover, it was suggested that,
due to their higher mobility at low temperature, only the electron
pocket is detected in quantum oscillations and no other frequency
has yet been observed.

Quantum oscillations are due to the oscillation of the density of
states of the quasi-particles in a magnetic field. The origin of
the Shubnikov-de Haas (SdH) effect (detected so far in
YBa$_2$Cu$_3$O$_{6.5}$), i.e. quantum oscillations of the
resistance, is basically due to an increase of scattering rate
which occurs when a Landau level crosses the Fermi level, inducing
a maximum in the resistance. Regarding the Haas-van Alphen (dHvA)
effect, i.e. quantum oscillations of the thermodynamic
magnetization, it is a direct consequence of the oscillation of
the free energy in magnetic field. In particular, it is not
sensitive to quantum interference effects \cite{Stark71}, which
can contaminate the electrical resistance.

In this letter, we report the first direct observation of dHvA
oscillations in YBa$_2$Cu$_3$O$_{6.5}$ at very low temperatures in
magnetic fields up to 59 T, using a piezoresistive cantilever. We
found a dHvA frequency of $F=540\pm4$~T and a cyclotron mass of
$m^*$=1.76 $\pm$ 0.07~m$_0$, in excellent agreement with the
values reported in the SdH measurements, i.e. $F_{SdH}=530\pm20$~T
and $m^*_{SdH}$=1.9 $\pm$0.1m$_0$. We also present the first study
of the angular dependence of the SdH frequency, confirming that
the oscillations arise from a quasi-2D Fermi surface. As all
aspects of the data obey the standard Lifshitz-Kosevich theory,
there is no indication of any deviation from Fermi-liquid theory.

\begin{figure}[t]
\centering
\includegraphics[width=1\linewidth,angle=0,clip]{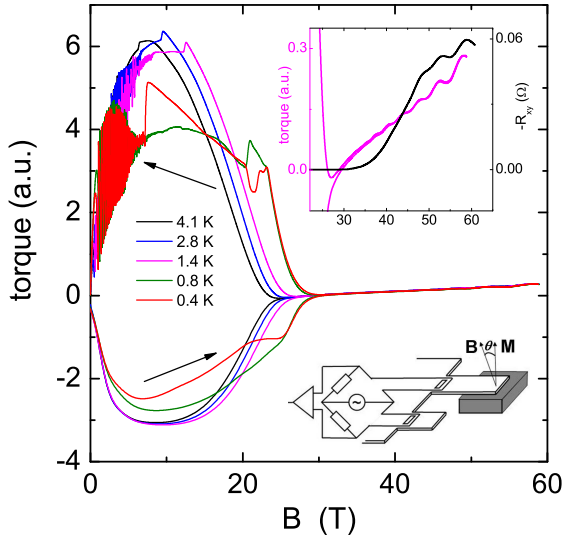}
\caption{(color on line) Torque versus magnetic field at different
temperatures. Both up (50~ms) and down (250~ms) field sweeps are
shown, as indicated by the arrows. Lower inset: schematic of the
cantilever. Upper inset: Magnetic fields dependence of the torque
($\theta \simeq 5^\circ$) at $T$=1.4~K and of the Hall resistance
($\theta=0^\circ$) measured in the same sample at $T$=1.5~K.}
\label{fig1}
\end{figure}

We used a detwinned single crystal of YBa$_2$Cu$_3$O$_{6.5}$
flux-grown in a non-reactive BaZrO$_3$ crucible. The dopant oxygen
atoms were ordered into an ortho-II superstructure of alternating
full and empty CuO chains, yielding a superconducting transition
temperature $T_c$=57.5~K \cite{Liang00}. The sample used for the
torque experiment is a small part (140*140*40 $\mu$m$^3$) of the
sample studied in ref. \cite{Doiron07}. The Hall effect of the
original piece was measured at different orientations of the
magnetic field with respect to the CuO$_2$ planes, in order to
study the angular dependence of the SdH frequency. Torque
measurements were performed with a commercial piezoresistive
microcantilever \cite{Ohmichi02} down to 0.4~K  at the LNCMP in
Toulouse, using pulsed magnetic fields up to 59~T. The sample was
glued with Araldite epoxy to the cantilever. A one-axis rotating
sample holder allowed the angle ($\theta\sim5~^\circ$) to be
varied between the normal to the CuO$_2$ planes and the magnetic
field at ambient temperature. The cantilever was set inside a
vacuum tight resin capsule filled at room temperature with $^4$He
gas to ensure thermalization of the sample. This capsule sits in
the $^3$He/$^4$He mixture of the dilution fridge. The temperature
gradient between the sample located at the end of the cantilever
and the thermometer located in the mixing chamber was estimated by
measuring the critical field of a known compound under the same
experimental conditions. The temperature gradient is about
0.2$\pm$0.1~K at 0.4~K and negligible above 1~K. This uncertainty
does not affect the value of the physical parameters extracted
from the data. The variation of the piezoresistance of the
cantilever is measured with a Wheatstone bridge with an AC
excitation at a frequency of 63~kHz (see lower inset of
Fig.~\ref{fig1}).


Raw data of torque
$\tau=|\overrightarrow{M}\times\overrightarrow{B}|$ versus
magnetic field are shown in Fig.~\ref{fig1} for the up and down
field sweeps between $T$=4.1~K and $T$=0.4~K. As observed in the
mixed state of other type-II superconductor at low
field,\cite{Mola01} there is a strong hysteresis effect, which
reflects the penetration (expulsion) of magnetic field into (out
of) the sample when the magnetic field increases (decreases). In
this regime, 'flux jumps' are clearly observed and are associated
with a thermal instability corresponding to a reorganization of
the magnetic flux as it enters (exits) the sample. \cite{Mola01,
Legrand93} The hysteresis disappears above $B_{irr}$ and the
torque becomes almost linear with magnetic field. $B_{irr}$ marks
the field where the electrical resistance starts to rise and
corresponds to the vortex liquid phase. The upper inset of
Fig.~\ref{fig1} shows the torque and the (normalized) Hall
resistance versus magnetic field measured in the same sample at $T
\simeq$1.5~K. Above a field of $\sim$30~T, clear dHvA oscillations
are observed. \cite{Shoenberg} We have checked that the amplitude
of oscillations is identical for the up and down field sweeps,
which shows that no detectable self-heating occurs in the sample
during the measurement.

In the following, we will discuss only the oscillatory part of
torque, obtained by subtracting a monotonic background from the
raw data. The amplitude of the first harmonic of the dHvA
oscillations is interpreted using the Lifshitz-Kosevich (LK)
theory for a two-dimensional Fermi liquid:
\begin{equation}
\label{eq:LK}
{\tau_{osc}}=BA\tan\theta\sin[2\pi(\frac{F}{B}-\gamma)]
\end{equation}
where $F$ is the oscillation frequency and $\gamma$ is a phase
factor. We neglect any contribution from magnetic breakdown, spin
damping or any additional damping coming from the effect of
superconductivity on the dHvA amplitude. The amplitude of the
fundamental frequency is given by $A \propto R_{T}R_{D}$ where
$R_{T}$ and $R_{D}$ are the thermal ($R_{T} = \alpha T
m^*/Bsinh[\alpha T m^{*}/B]$) and Dingle ($R_{D} = exp[-\alpha T_D
m{^*}/B]$) damping factors, respectively, where $\alpha =
2\pi^2k_Bm_0/e\hbar$ ($\simeq$ 14.69~T/K) and $T_D$ is the Dingle
temperature. \cite{Shoenberg}


\begin{figure}[t]
\centering
\includegraphics[width=1\linewidth,angle=0,clip]{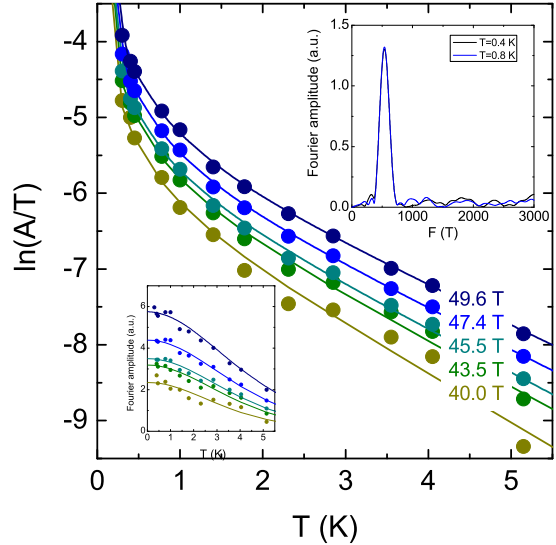}
\caption{(color on line) Main panel: Temperature dependence of the
oscillation amplitude A (where A is determined from discrete
Fourier analysis with a Blackmann window) plotted as ln(A/T)
versus T for various mean magnetic field values. Solid lines are
fits to Eq.\ref{eq:LK} with m$^*$ = 1.76~m$_0$. Lower Inset: same
data plotted as A versus temperature. Upper Inset: Fourier
Transform of the oscillatory part of the torque at $T$=0.4~K and
$T$=0.8~K between 35~T and 58~T, revealing a single frequency
$F$=540~T.} \label{fig2}
\end{figure}


Fig.~\ref{fig2} displays the temperature dependence of the
amplitude of the oscillations ("mass plot") for different field
windows. The usual behavior expected for thermal damping of the
dHvA oscillations is observed. Solid lines are fits of
Eq.~\ref{eq:LK}, from which we deduce an effective mass $m^*$ =
1.76 $\pm$ 0.07~$m_0$, in agreement with the value obtained from
Shubnikov-de Haas measurements \cite{Doiron07}. No significant
variation of the effective mass with magnetic field is observed,
which confirms that the LK model is adequate to describe the data.
Using $v_F=\hbar k_F / m^*$ and the Onsager relation (assuming a
circular orbit) $\pi k_F^2 = 2 \pi^2 F/ \Phi_0$, the average Fermi
velocity is $8.4\times 10^4$~m~s$^{-1}$. This value is in fact
very similar to that found by ARPES at the anti-node in both
La$_{2-x}$Sr$_x$CuO$_4$ \cite{Yoshida07} and
Bi$_2$Sr$_2$CuO$_{6+\delta}$. \cite{Kaminski05} This may indicate
that the pockets detected by dHvA are not located at the same
position in $k$-space as the Fermi arcs seen by ARPES
\cite{Norman98}.


\begin{figure}[t]
\centering
\includegraphics[width=1\linewidth,angle=0,clip]{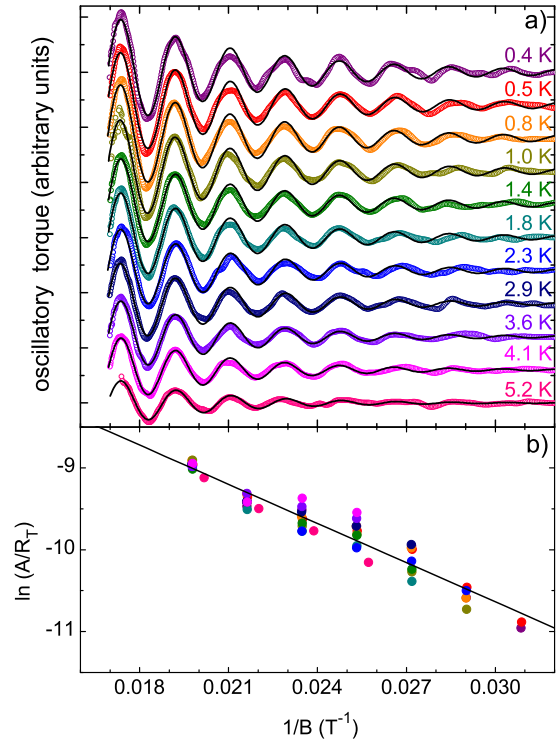}
\caption{(color on line) (a) Field dependence of the oscillatory
part of the torque for the various temperatures explored and (b)
corresponding Dingle plots. Solid lines are best fits to
Eq.\ref{eq:LK} with $m^*$ = 1.76~$m_0$ and T$_D$ = 6.2 K.}
\label{fig3}
\end{figure}


Fig.~\ref{fig3}(a) displays the oscillatory torque versus 1/$B$
between $T$=5.2~K and $T$=0.4~K. At the lowest temperatures, 8
periods can be resolved, corresponding to the Landau levels $n$=9
to $n$=16. Oscillations can be observed down to a field of
$\sim$30~T, in particular in the resisitive superconducting
transition (see upper panel of Fig.~\ref{fig1}).
Fig.~\ref{fig3}(b) displays the corresponding field dependence of
the amplitude of the dHvA oscillations divided by the $R_T$ factor
("Dingle plot"). Within our experimental resolution, it is not
possible to resolve an extra attenuation corresponding to the
effect of superconductivity on the dHvA amplitude.
\cite{Janssen98, Fletcher04} By neglecting any additional damping
factor coming from the mixed state, the solid line in
Fig.~\ref{fig3}(b) yields an upper limit of $T_D$ = 6.2 $\pm$
1.2~K (using $m^*$=1.76 $\pm$0.07~$m_0$), which converts to
$\omega_c\tau$ = 0.7 $\pm$ 0.2 at $B$=35~T. Assuming a cylindrical
FS, we can deduce a mean free path $l$=16~nm. Black solid lines in
Fig.~\ref{fig3}(a) are fits to Eq.~\ref{eq:LK} obtained by setting
$m^*$ = 1.76~$m_0$ and $T_D$ = 6.2~K for all temperatures. The
deduced oscillation frequency and phase factor are $F$ = 540 $\pm$
4~T and $\gamma$ = 0.15 $\pm$~0.05, respectively. This fitting
procedure confirms that the LK formula, which describes the dHvA
oscillations for a 2D Fermi liquid is appropriate. This conclusion
is at odds with recent theories that invoke mechanisms other than
the Landau quantization of quasi-particles. \cite{Alexandrov07,
Melikyan07}

\begin{figure}[t]
\centering
\includegraphics[width=1\linewidth,angle=0,clip]{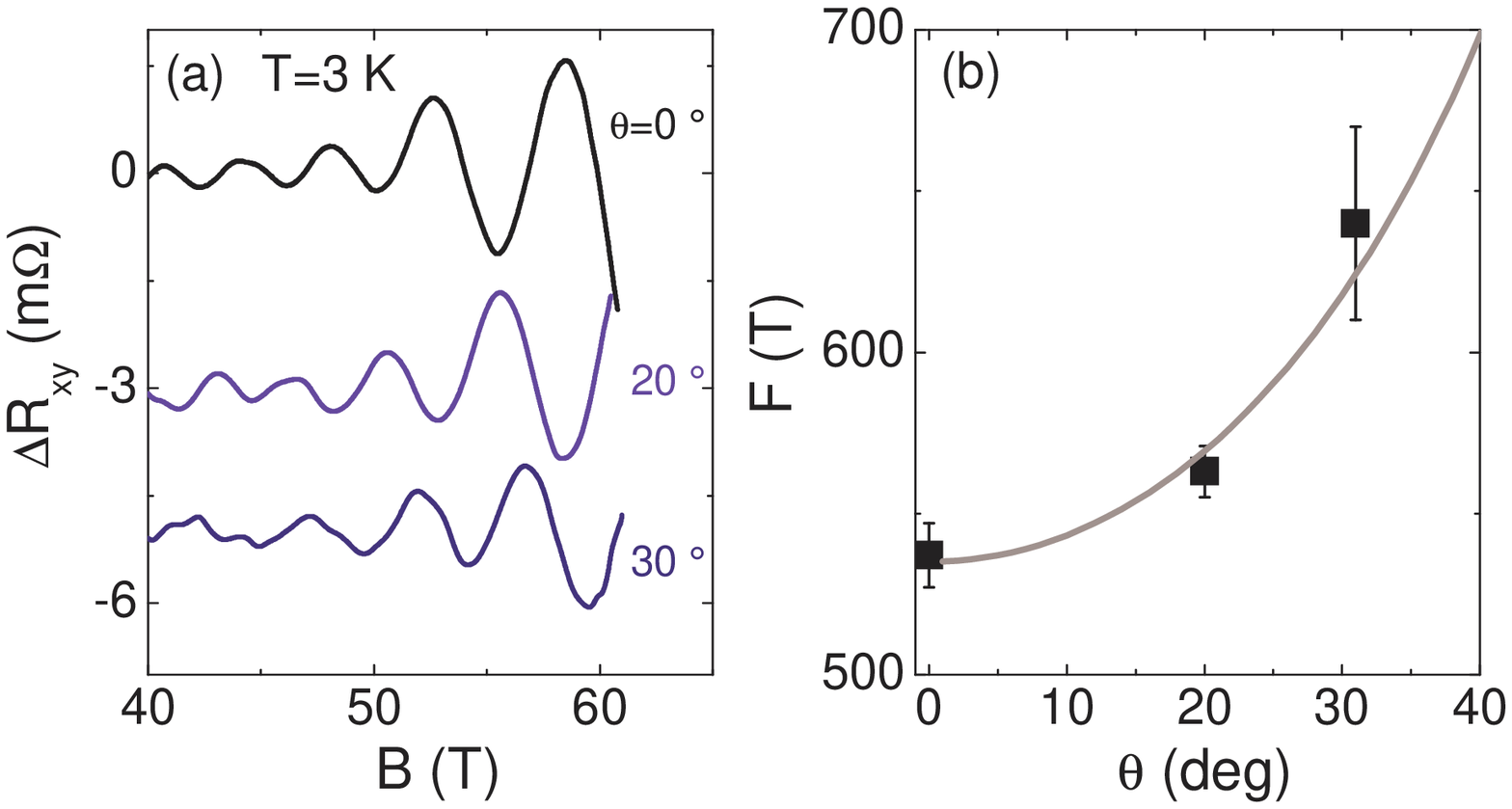}
\caption{(color on line) (a) Oscillatory part of the Hall
resistance versus magnetic field at different angles at T=3~K. (b)
Angular dependence of the deduced Shubnikov de Haas frequency.
Solid line is a fit to F($\theta$=0)/cos($\theta$). } \label{fig4}
\end{figure}

Fig.~\ref{fig4}(a) shows the oscillatory Hall resistance versus
magnetic field measured in the same sample at $T$=3~K for
different angles $\theta$. Several measurements have been done at
different temperatures for each orientation, from which we have
deduced an oscillation frequency $F$ plotted in
Fig.~\ref{fig4}(b). Within the error bars, $F$ increases as
1/cos($\theta$) as expected for a quasi-2D part of the FS
\cite{Shoenberg}. It simply reflects the increase of the area of
the cyclotron orbit when the magnetic field is tilted from the
normal of the CuO$_2$ plane.


dHvA measurements have been widely used to study the FS of metals.
\cite{Shoenberg} As a thermodynamic measurement, it firmly
confirms the existence of well-defined quasi-particles at the FS
with a substantial mean free path. Quantum oscillations are
detected with torque starting at a field corresponding to the
onset of the electrical resistance, but there is no clear evidence
in the amplitude of the oscillations of a transition from the
mixed state to the normal state. \cite{Yasui02} It may be due to
the fact that, in order to resolve an additional damping, the dHvA
signal have to be measured in a field range much larger than than
in Fig.~\ref{fig3}(b). dHvA oscillations have been observed in the
mixed state of many superconductors, \cite{Graebner76, Mueller92,
Bergemann97, Janssen98, Clayton02, Fletcher04} and in all cases
the oscillations have the same frequency as in the normal state.
In particular, dHvA and SdH effects have been observed in the
organic superconductor $\kappa$-(BEDT-TTF)$_2$Cu(NCS)$_2$
\cite{Sasaki03} which shares many characteristics with high $T_c$
superconductors such as low dimensionality and short coherence
length, but which has a much lower $H_{c2}$. \cite{Belin99,
Sasaki98, Lortz07} The striking similarity between the data in
$\kappa$-salt (see Fig.~2 of ref.~\cite{Sasaki03}) and that in the
upper inset of Fig.~\ref{fig1} suggests that the normal state
could be reached around our maximum field in
YBa$_2$Cu$_3$O$_{6.5}$. This value is lower than the one given by
the extrapolation of the Nernst data to higher fields,
\cite{Wang03} which suggest that $H_{c2}$ could be as large as
150~T (corresponding to the field scale associated with the short
coherence length in cuprates). However, one should keep in mind
that in such a layered superconductor, there is no true phase
transition between the vortex liquid and the normal state, but a
cross-over with an extended range of fluctuating
superconductivity. \cite{Rullier07} In this cross-over regime, one
can expect the electrical resistance to be representative of the normal state.\\
Finally, the frequency $F$=540~T converts to a carrier density
0.038 carrier per planar Cu atom. Independently of whether there
are 2 or 4 pockets, it gives a number of carriers which is not in
agreement with the doping level (10~$\%$). However, a scenario
based on a ($\pi ,\pi$) reconstruction of the FS can explain both
the negative Hall effect (electron pocket) in the normal state
\cite{LeBoeuf07} and the apparent violation of the Luttinger sum
rule. It assumes that the frequency observed with SdH and dHvA
effects corresponds to an electron pocket, whose mobility is much
higher than that of a larger hole pocket not seen in the
measurements. To validate this scenario, however, the detection of
another frequency corresponding to a total hole density of 0.138
hole per planar Cu atom, is still required.


In summary, we have observed the first direct evidence of dHvA
oscillations in YBa$_2$Cu$_3$O$_{6.5}$. The frequency and the
effective mass are in excellent agreement with previous SdH
measurements in the same compound. This observation confirms the
existence of a coherent closed Fermi surface at low temperature in
the underdoped side of the phase diagram of cuprates and suggests
that the ground state obeys Fermi-liquid theory.

We thank K. Behnia, N. Hussey and G. Rikken for useful
discussions. We acknowledge support from EuroMagNET under EC
contract 506239, the Canadian Institute for Advanced Research, the
ANR project ICENET and funding from NSERC, FQRNT, EPSRC, and a
Canada Research Chair.


\end{document}